\documentclass[conference]{IEEEtran}

\usepackage{cite}
\usepackage{amsmath,amssymb,amsfonts}
\usepackage{algorithmic}
\usepackage{graphicx}
\usepackage{tabularx}
\usepackage{textcomp}
\usepackage{xcolor}
\usepackage{tcolorbox}
\usepackage{hyperref}
\usepackage{placeins}
\usepackage{multirow}
\usepackage{balance}
\graphicspath{ {figures/} }

\usepackage{eso-pic}

\begin{document}
\bstctlcite{IEEE-NoDash:BSTcontrol} 

\title{Faster Releases, Fewer Risks: A Study on Maven Artifact Vulnerabilities and Lifecycle Management}

\author{\IEEEauthorblockN{
Md Shafiullah Shafin$^1$  ~~~~~~~~ Md Fazle Rabbi$^2$ ~~~~~~~~ S. M. Mahedy Hasan$^1$ ~~~~~~~~ Minhaz F. Zibran$^2$}
\IEEEauthorblockA{
\textit{$^1$Department of Computer Science, Rajshahi University of Engineering \& Technology, Rajshahi, Bangladesh} \\
\textit{$^2$Department of Computer Science, Idaho State University, Pocatello, ID, United States} \\
shafiullahshafin735@gmail.com, mdfazlerabbi@isu.edu, mahedy@cse.ruet.ac.bd, zibran@isu.edu}
}


\maketitle

\begin{abstract}
In modern software ecosystems, dependency management plays a critical role in ensuring secure and maintainable applications. However, understanding the relationship between release practices and their impact on vulnerabilities and update cycles remains a challenge. In this study, we analyze the release histories of 10,000 Maven artifacts, covering over 203,000 releases and 1.7 million dependencies. We evaluate how release speed affects software security and lifecycle.
Our results show an inverse relationship between release speed and dependency outdatedness. Artifacts with more frequent releases maintain significantly shorter outdated times. We also find that faster release cycles are linked to fewer CVEs in dependency chains, indicating a strong negative correlation. These findings emphasize the importance of accelerated release strategies in reducing security risks and ensuring timely updates. Our research provides valuable insights for software developers, maintainers, and ecosystem managers.

\end{abstract}


\section{Introduction}
\label{sec:intro}

In modern software engineering, reusing existing software components is key to improving development efficiency and maintaining software quality~\cite{qian2023communicative}. Software reuse accelerates development cycles and reduces costs~\cite{segun2024developing}. This allows developers to focus on innovation instead of reinventing the wheel. Understanding the dynamics of software ecosystems is crucial for ensuring the sustainability and resilience of software systems.


Among the ecosystems that support software reuse, Maven stands out as one of the most widely used. Maven is a build automation tool that helps manage dependencies and automate project lifecycles. It facilitates the integration of third-party libraries~\cite{salvi2024demystifying}. The Maven ecosystem consists of a large network of artifacts, which are different versions of software components that often serve as dependencies for other artifacts. Regular updates to these artifacts introduce new features, fix bugs, and address security vulnerabilities, all of which contribute to the health of the software ecosystem.


Despite its advantages, the Maven ecosystem faces significant challenges. One of the main issues is managing outdated dependencies, which occur when artifacts rely on older versions of dependencies. Outdated dependencies lead to vulnerabilities, performance issues, and compatibility problems. Another critical concern is the presence of Common Vulnerabilities and Exposures (CVEs), which are publicly disclosed security flaws that can be exploited, posing a serious threat to software security.

Over the years, researchers have proposed various tools and methods to address these challenges, such as automated dependency updates and security scanners \cite{Valero2021}\cite{Liu2022}\cite{Pashchenko}. Some of them have conducted studies on the Maven packages to see the effect of transitivity and granularity on vulnerability propagation in the ecosystem \cite{Mir}.  However, many existing solutions do not fully explore how quickly artifacts release and how this affects the risks of outdated dependencies and CVEs.

This research fills a gap by addressing two main research questions (RQs): 

\vspace{0.1cm}
\textit{\textbf{\emph{RQ1:}} How does the speed of releasing new artifact versions affect how outdated their dependencies become?}  

\vspace{0.1cm}
\textit{\textbf{\emph{RQ2:}} How does the speed of artifact releases impact the number of security vulnerabilities in their dependencies?}  
\vspace{0.1cm}

To address these questions, we analyze a subset of the Maven dependency graph database~\cite{Jaime2025} containing 10,000 Maven artifacts, with 203,861 release versions and 1,732,229 dependency releases. For each dependency release, we calculate how long it remains outdated and identify any associated CVEs.  This helps us understand the relationship between release speed, dependency freshness, and vulnerability propagation. Additionally, we examine the average release frequency of artifacts and its impact on ecosystem health.
This analysis offers insights into the propagation of outdated dependencies and vulnerabilities concerning the average number of releases of an artifact in the Maven ecosystem. For replication purposes, we publicly release a replication package~\cite{replicationPackage} including the queries, scripts and data used in this study.

The subsequent sections start with a description of the dataset in Section~\ref{sec:dataset-description}, followed by an explanation of the research methodology in Section~\ref{sec:research-methodology} and detailed analysis and findings in Section~\ref{sec:analysis-findings}. The threats to validity are discussed in Section~\ref{sec:threats-validity}, while Section~\ref{sec:related-work} covers a discussion on related work. The exploration concludes in Section~\ref{sec:conclusion}.

\section{Dataset Description}
\label{sec:dataset-description}

To conduct this study on the Maven ecosystem, we extract data from a Neo4j dump file titled \texttt{with\_metrics\_goblin\_maven\_30\_08\_24.dump}\cite{Jaime2025}, which contains the entire Maven Central dependency graph. This dataset consists of a total of 59,152,712 different types of nodes.

One type of node in the graph is \texttt{Artifact}, representing a component such as a library, framework, or tool. Another type of node is \texttt{Release}, which refers to a specific version of an artifact that is officially published and made available for use by other developers and projects. The release node contains the version name and the release timestamp in Unix format. Artifact nodes connect to their corresponding release nodes through a one-to-many relationship.

There is also a \texttt{Dependency} edge that connects release nodes to other artifact nodes through a many-to-many relationship. This dependency edge contains two values:  
\begin{itemize}  
\item \texttt{scope}: This refers to the stage of the development lifecycle in which the release depends on a specific version of another artifact.  
\item \texttt{targetVersion}: This is the specific version of the artifact that the dependent release relies on.  
\end{itemize}  


Additionally, there is another set of nodes, \texttt{AddedValue}, classified into four types: \texttt{SPEED}, \texttt{FRESHNESS}, \texttt{CVE}, and \texttt{POPULARITY\_1\_YEAR}. This study focuses on SPEED, FRESHNESS, and CVE.  
\begin{itemize}  
\item \texttt{SPEED}: This node contains a fractional value representing the rate at which new versions of an artifact release.  
\item \texttt{FRESHNESS}: This refers to how recently a release updates or releases. This node contains two values:  
    \begin{itemize}  
    \item \texttt{numberMissedRelease}: The number of new versions released after the given version.  
    \item \texttt{outdatedTimeInMS}: A numeric timestamp indicating how old or stale the version is compared to the latest available version.  
    \end{itemize}  
\item \texttt{CVE}: This node contains information about known vulnerabilities related to a specific artifact version.  
\end{itemize}  

AddedValue nodes connect to release nodes through a one-to-one relationship. The speed-type AddedValue node connects to its corresponding artifact through an edge. Each type of node has a unique string-type identifier. 



\section{Methodology}
\label{sec:research-methodology}

Initially, we run a query (Query-A) on the Neo4j desktop application to randomly select 10,000 artifacts for our study. We export all these artifact IDs into a CSV file. Next, we load this CSV file into the application and run another query (Query-B) on these artifact IDs to retrieve all the releases of each artifact. We export the release IDs along with the artifact IDs into a new CSV file. The total number of releases is 203,861.

We then load the latest CSV file into the application and run a query (Query-C) to find all the direct dependency releases for each release in the second CSV file. After retrieving all the direct dependency releases, we export them along with the dependent releases into another CSV file. The total number of dependency releases is 1,732,229. 

Next, we load this CSV file into the application and perform two additional queries (Query-D and Query-E) on all dependency releases to retrieve the outdatedTimeInMS and numberOfCVEs for each dependency release, respectively. We extract another CSV file containing this dependency-release information, specifically outdatedTimeInMS and numberOfCVEs.

\begin{figure}[htbp]
    \hspace{-0.3cm}
    \includegraphics[width=0.52\textwidth]{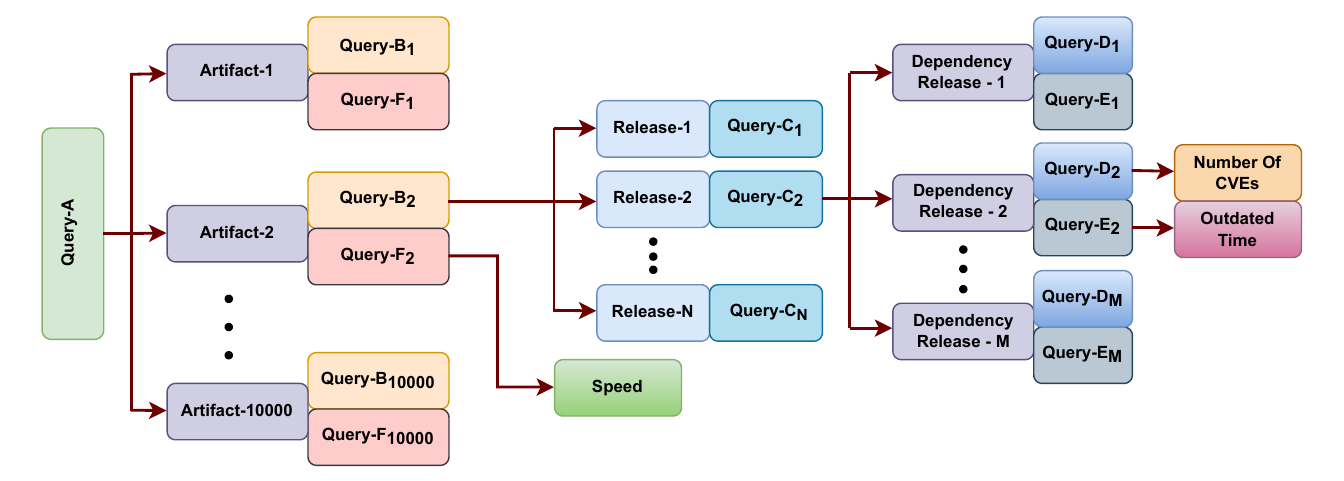}
    \vspace{-0.5cm}
    \caption{Process for selecting artifacts and their detailed release dependencies}
    \label{fig:workflow}
\end{figure}

Finally, we load the first CSV file, which contains only artifact IDs, and run another query (Query-F) to extract the releaseSpeed value for each artifact. Figure~\ref{fig:workflow} illustrates the data extraction process, showing the selection of 10,000 artifacts, their corresponding releases, and their associated dependency releases.

\textbf{Determining average outdatedTimeInDays of an artifact:}  
After extracting the \textit{outdatedTimeInMS} for each dependency release, we convert the value to \textit{outdatedTimeInDays} for computational convenience.  

Next, we average all the \textit{outdatedTimeInDays} values from the dependency releases associated with the same release. This average \textit{outdatedTimeInDays} is then assigned to the corresponding release. Similarly, we average the \textit{outdatedTimeInDays} values for each release based on the same artifact, obtaining the average \textit{outdatedTimeInDays} for each artifact. This averaged value is then assigned to the corresponding artifact.

\textbf{Determining average numberOfCVEs of an artifact:}  
Similarly, after extracting \textit{numberOfCVEs} for each dependency release, we average the \textit{numberOfCVEs} values from the dependency releases associated with the same release. This average \textit{numberOfCVEs} is then assigned to the corresponding release. Again, we average the \textit{numberOfCVEs} values for each release based on the same artifact, obtaining the average \textit{numberOfCVEs} for each artifact. This averaged value is then assigned to the corresponding artifact.

\textbf{Determining releaseSpeed of an artifact:}  
For each artifact, we obtain its \textit{releaseSpeed} as a fractional value using a single query. We then convert this value to the \textit{releaseSpeed} per year. 

In this way, we determine the average \textit{numberOfCVEs}, \textit{outdatedTimeInDays}, and \textit{releaseSpeed} for each artifact. This approach allows us to conduct an artifact-centered study on the data.  

To better understand the relationships between release speed, outdatedness, and security vulnerabilities, we perform Pearson's correlation analysis on the datasets. Pearson's correlation coefficient ($r$) measures the linear relationship between two variables and ranges from -1 (perfect negative correlation) to +1 (perfect positive correlation). The formula for Pearson’s correlation is:

\begin{equation}
r = \frac{\sum (x_i - \bar{x})(y_i - \bar{y})}{\sqrt{\sum (x_i - \bar{x})^2 \sum (y_i - \bar{y})^2}}
\end{equation}

\vspace{-0.2cm}
Where:
\begin{itemize}
    \item $x_i$ and $y_i$ are the individual data points.
    \item $\bar{x}$ and $\bar{y}$ are the means of the respective datasets.
\end{itemize}

\vspace{0.1cm}
\section{Analysis And Findings}
\label{sec:analysis-findings}

\subsection{Impact of Release Speed on Dependency Freshness (RQ1)}

\begin{figure}[htbp]
    \centering
    \includegraphics[width=\columnwidth]{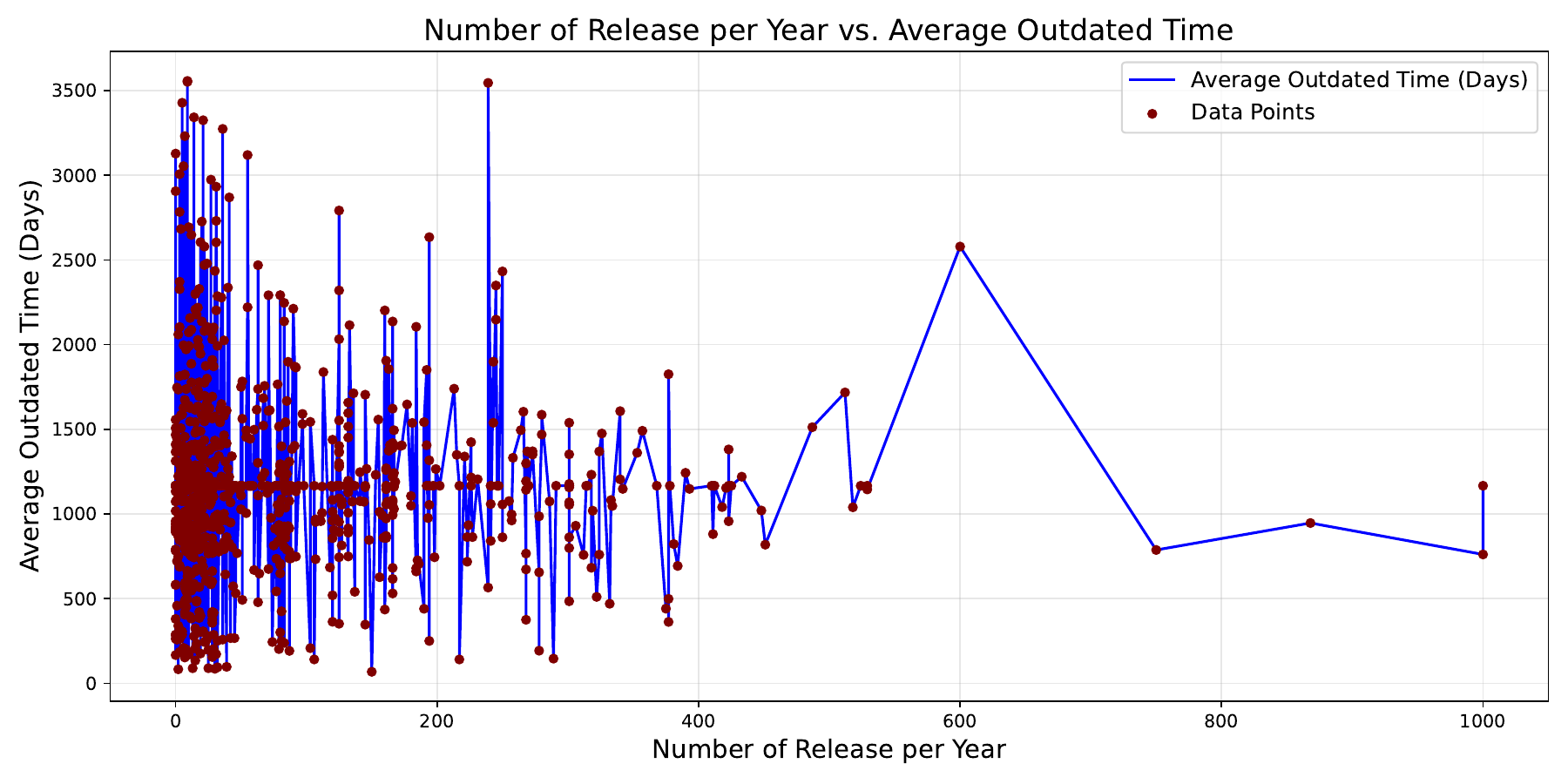}
    \vspace{-0.3cm}
    \caption{Relationship between release speed and dependency freshness}
    \label{fig:speedvsoutdated}
\end{figure}

We investigate the relationship between release speed and dependency freshness using data from the Maven Dependency Graph Database. Our analysis focuses on the average outdated time for dependencies based on artifact release speed. Figure~\ref{fig:speedvsoutdated} shows the number of releases per year on the x-axis and the average \textit{outdatedTime} (in days) on the y-axis. The figure reveals an inverse relationship, indicating that faster release speeds correlate with fresher dependencies.

Artifacts with higher release frequencies, especially those releasing more than ten times per year, have shorter outdated times, typically under 500 days. In contrast, artifacts with fewer than five releases annually tend to have outdated times over 1,500 days, with some reaching almost 3,000 days. This trend highlights the impact of release frequency on how up-to-date dependencies are. Faster-releasing artifacts are more likely to adopt newer dependency versions, reducing exposure to risks like vulnerabilities, performance issues, and compatibility problems. Slower release cycles, however, result in outdated dependencies that affect not just the artifact but other projects relying on it.

The correlation coefficient of 
$r = -0.4211$ shows a moderate negative relationship between release speed and outdated time. This suggests that while the correlation isn't very strong, frequent updates play an important role in keeping dependencies fresh. The data indicates that projects with faster release cycles can quickly integrate improvements and fixes, which strengthens the ecosystem. This is especially important in fast-evolving fields where delayed updates lead to technical debt, compatibility issues, and missed optimization opportunities.

\begin{tcolorbox}[boxrule=0.5pt, boxsep=-2pt, left=5pt, right=5pt]
    \textbf{Ans. to RQ1:} Artifacts with faster release speeds have fresher dependencies. 
        Faster release cycles help maintain updated versions and reduce outdatedness.
\end{tcolorbox}


\subsection{Effect of Release Speed on Vulnerability Frequency (RQ2)}
\begin{figure}[htbp]
    \centering
    \includegraphics[width=\columnwidth]{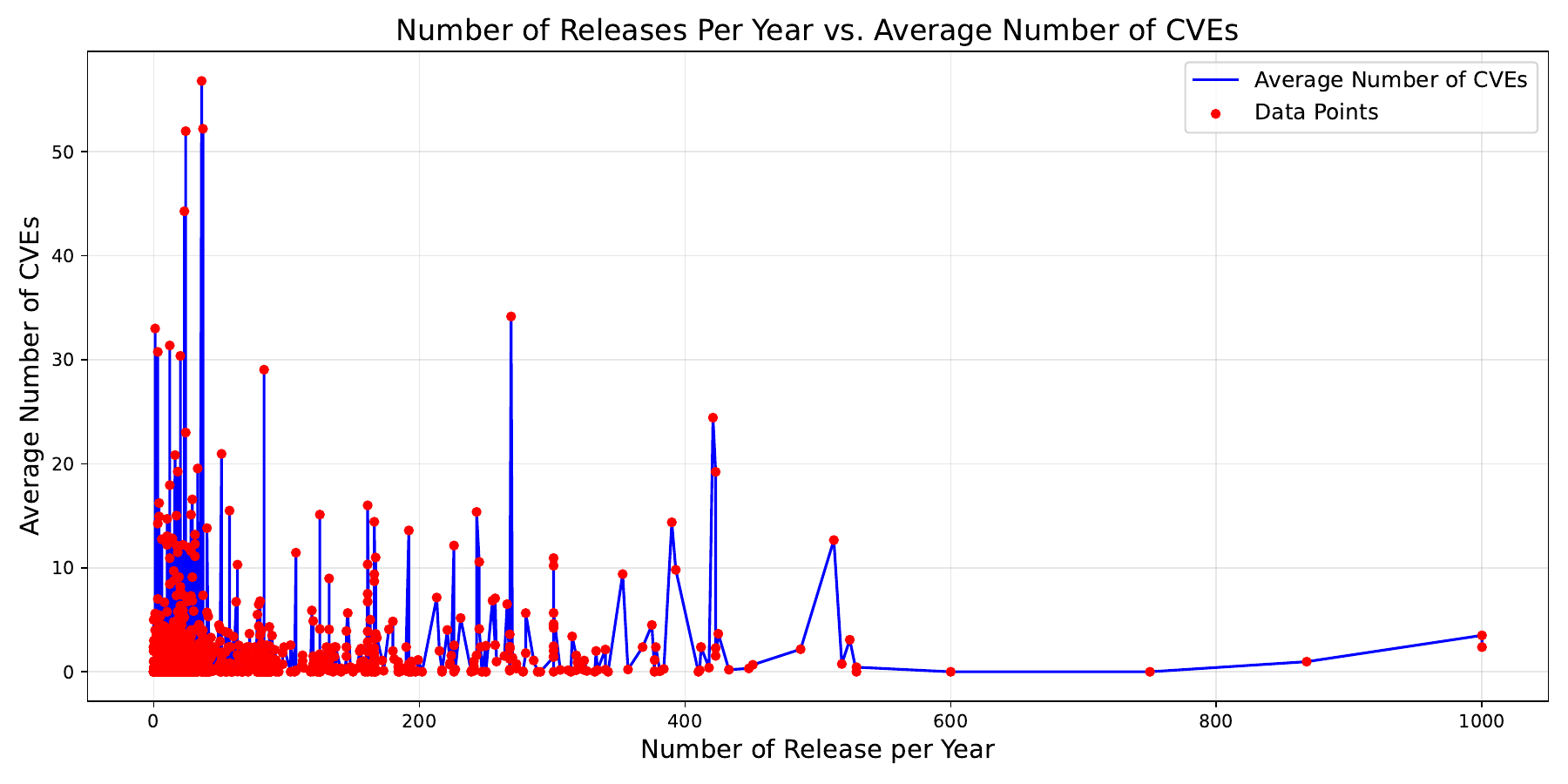}
    \vspace{-0.3cm}
    \caption{Release speed and its impact on vulnerabilities in dependencies}
    \label{fig:speedvscve}
\end{figure}

We explore the relationship between artifact release speed and the frequency of security vulnerabilities in their dependencies by analyzing the number of CVEs associated with artifacts across different release speeds. Figure~\ref{fig:speedvscve} shows this relationship, with the number of releases per year on the x-axis and the average number of CVEs on the y-axis. The figure clearly shows that artifacts with faster release cadences tend to have fewer CVEs in their dependency chains, indicating a strong link between frequent updates and better security.


Artifacts releasing fewer than two versions per year have an average of 20 CVEs per release, with some data points approaching or exceeding 30 CVEs. On the other hand, artifacts with release rates above 10 versions per year consistently have fewer than 5 CVEs on average. This trend highlights the importance of frequent updates in reducing exposure to known vulnerabilities. Regular releases help projects adopt patched and secure versions of dependencies more quickly, reducing the time vulnerabilities remain exploitable. In contrast, slower release cycles result in prolonged use of outdated and vulnerable dependencies, increasing the risk of security breaches.

The negative correlation between release speed and the number of CVEs, 
$r = -0.4977$, shows the effectiveness of frequent releases in reducing security risks. This correlation is stronger than the one found in RQ1, indicating that release speed plays a more significant role in improving security than in maintaining dependency freshness. Faster release cycles allow projects to stay ahead of emerging security threats, addressing vulnerabilities as they are discovered and fixed.

\begin{tcolorbox}[boxrule=0.5pt, boxsep=-2pt, left=5pt, right=5pt]
        \textbf{Ans. to RQ2:} Artifacts with faster release speeds tend to have fewer CVEs in their dependencies. 
        Frequent updates allow projects to adopt secure versions, reducing vulnerability exposure.
\end{tcolorbox}

\section{Threats to validity}
\label{sec:threats-validity} 

While our research offers valuable insights, several potential threats impact the robustness of our findings. First, our analysis relies on 10,000 artifacts, which, though diverse, may not fully represent the entire Maven ecosystem, potentially overlooking the behavior of niche or less frequently used artifacts. Sampling bias may also arise, as highly popular artifacts with frequent updates dominate the dataset, skewing results. 

Additionally, the dataset reflects a specific point in time, and evolving release practices or vulnerability disclosures affect the applicability of our conclusions in the future. The reliance on publicly available CVE data poses another challenge, as undisclosed or newly discovered vulnerabilities may lead to an underestimation of security risks. 

Furthermore, we treat all dependencies uniformly without accounting for their specific usage contexts (e.g., runtime or testing), which could influence the observed correlations. Assumptions made in calculating release speed, particularly for artifacts with irregular update patterns, introduce noise into our analysis. Lastly, external factors such as organizational policies, resource constraints, and developer practices, which are beyond our scope, also influence release and dependency management behaviors. Addressing these limitations in future research could further validate and enhance the generalizability of our findings.

\section{Related Work}
\label{sec:related-work}
There have been many studies on software quality assurance that involve the investigation of bug patterns~\cite{Islam_BugPattern_SAC2020,Islam_BugPattern_ACR2021,Amit_ChromiumBug_2022}, 
vulnerabilities~\cite{Islam_2016_IWSC,Islam_2017_ESEM}, code smells~\cite{Islam_IWSC_2018,Zibran_2013_RefactorSchedule,Zibran_2013_CloneChangeStudy}, code quality~\cite{Duaa_2018_ComplexityReadability,Islam_IWSC_2018}, human aspects~\cite{Rabbi_2023_PhoneSensor,Champa_2023_Female,Islam_2016_IJSI,Islam_2016_SERA,Islam_BugSentiment_SEDE2018} of software development and maintenance as well as comparison of methods/tools~\cite{Islam_2017_DictionaryConstruction,Islam_2018_SentiToolCompare,Ryan_AndroidSec_2021,Daniel_WordPress_2021} for measuring such aspects.

Several studies have been performed on the Maven ecosystem, analyzing artifact release vulnerabilities, popularity, release speed, and freshness. Keshani et al. \cite{Keshani} observed that popularity does not play a role in whether or not a method is involved in breaking changes. Ma et al. \cite{Ma2024} proposed a tool named VulNet to address the limitations found in Maven dependency regarding a lack of prioritization mechanisms on which dependencies are more likely to cause an issue. Mir et al. \cite{Mir} conducted a study on 3M Maven packages to see the effect of transitivity and granularity on vulnerability propagation in the ecosystem. Yulun et al. \cite{Yulun2023} conducted an analytical study to understand the potential threats of upstream vulnerabilities to downstream projects in the Maven ecosystem. Zhang et al. \cite{Zhang} proposed a solution for range restoration (Ranger) to automatically restore the compatible and secure version ranges of dependencies for downstream dependents. Valero et al. \cite{Valero2021} propose DepClean, a tool to determine the presence of bloated dependencies in Maven artifacts. Bloated dependencies are libraries that are packaged with the application’s compiled code but that are actually not necessary to build and run the application. Shen et al. \cite{Shen2024} conducted an study on vulnerability source and the fine-grained vulnerability propagation, localization, and repair of libraries and their corresponding client programs. 

Alfadel et al. \cite{Alfadel2023} conducted a study on the PyPi ecosystem to observe the propagation and life span of security vulnerabilities, accounting for how long they take to be discovered and fixed. Liu et al. \cite{Liu2022} proposed a knowledge graph-based dependency resolution, which resolves the inner dependency relations of dependencies as trees (i.e., dependency trees), and investigates the security threats from vulnerabilities in dependency trees at a large scale.
Bodin et al. \cite{Chinthanet2021} tried to identify and characterize vulnerability in its propagation and how to mitigate propagation lags in an ecosystem. Pashchenko et al. \cite{Pashchenko} proposed Vuln4Real, the methodology for counting actually vulnerable dependencies, that addresses the over-inflation problem of academic and industrial approaches for reporting vulnerable dependencies. Wang et al. \cite{Wang} conducted an empirical study to learn the scale of packages that block the propagation of vulnerability fixes in the ecosystem and investigate their evolution characteristics. D Jaime et al. \cite{DJaime} provided an optimal update strategy that aligns with developer priorities and minimizes incompatibilities. 

For JavaScript vulnerability detection, Ferreira et al.\cite{Ferreira} introduced the Multiversion Dependency Graph (MDG), a new graph-based data structure that captures the evolution of objects and their properties during program execution. Imranur et al.\cite{rahman} compared package update speeds across ecosystems, finding that PyPi packages update dependencies faster than NPM and Cargo. Finally, Runzhi et al.~\cite{Runzhi} summarized the key characteristics of an ideal dependency management bot, focusing on configurability, autonomy, transparency, and self-adaptability.

\section{Conclusion}
\label{sec:conclusion}

This study highlights the importance of release speed in managing dependency freshness and security within the Maven ecosystem. We analyze 10,000 artifacts and find that artifacts with faster release cycles have fewer outdated dependencies and fewer security vulnerabilities. Artifacts that release updates more frequently show lower outdated times and fewer CVEs. This demonstrates that frequent updates not only improve the relevance of dependencies but also enhance software security.
Our statistical analysis, including Pearson’s correlation coefficients, supports this relationship. We find a moderate to strong negative correlation between release speed and both outdated time and CVEs. These results show the importance of faster release cycles in keeping dependencies up-to-date and reducing exposure to known vulnerabilities. This is crucial in the constantly changing software landscape.

We suggest adopting practices like continuous integration and deployment (CI/CD) to ensure artifacts remain current and secure. These practices can enable faster release cycles. These insights offer a foundation for improving release strategies in software development, especially in large ecosystems like Maven, where dependencies are shared across many projects. However, our study relies on a subset of data and excludes certain factors, such as usage context and external influences, which may limit the generalizability of our findings. Future research can address these limitations by incorporating a broader dataset and considering additional factors like the context of dependency use.
Overall, the study shows the value of maintaining rapid release cycles to improve software security and keep dependencies relevant, contributing to more stable and secure software projects in the Maven ecosystem.

\balance

\section*{Acknowledgement}
This work is supported in part by the ISU-CAES (Center for Advanced Energy Studies) Seed Grant at the Idaho State University, USA.

\bibliographystyle{IEEEtran}
\bibliography{ref,Bug,SBOM_JS,sentiment,phishing,gender}

\end{document}